# Constructing an Artworld Influencers Network by Mining Social Media

Amalia Foka

# Constructing an Artworld Influencers Network by Mining Social Media

Amalia Foka (researcher and educator), University of Ioannina, Department of Fine Arts & Art Sciences, P.O. Box 1186, 45110, Ioannina, Greece. Email afoka@uoi.gr. ORCID: 0000-0001-7776-1308

## <1>ABSTRACT

The author shows a methodology for constructing a network map of the artworld. The tweets posted by a manually comprised set of exemplar artworld actors and Natural Language Processing (NLP) methods are used to identify significant artworld actors, including those that do not own a Twitter account. Identified artworld actors form communities detected with the Louvain method, and their authority is evaluated with the HITS centrality method. The resulting artworld network exhibits the ability, starting with a small set of exemplar actors, to identify new artworld actors and provides a wealth of data that would be very hard to collect otherwise.

<1>PROBLEM

The artworld is regarded as a network of actors with various roles and positions that interact and determine the cultural value of artworks [1]. Although at different periods of time, the notion of who constitutes the artworld and how they interact might vary, what remains constant is that an artwork's cultural value is primarily socially constructed by those participating in the artworld. Moreover, actors in the artworld compete to gain more power and influence the social construction of value for artworks [2]. This aspect is also reflected in the various published art-related rankings [3].

If we depict the artworld as a network, where the artworld actors constitute its nodes that are connected with edges to denote interactions amongst them, then it is in constant flux in terms of its structure as well as in terms of each actor's power and influence over others. However, it is not an easy task to identify all artworld actors nor to depict how they influence each other.

This paper initiates a discussion on how mining social media data can be used to analyze the artworld. Social media users engage in public conversations and post their opinion about a great variety of subjects. Social media users' messages and interactions provide a wealth of information that would otherwise be obtained by cumbersome procedures like surveys and interviews. Consequently, the continuous increase of social media users and content has given rise to numerous methodologies that have been proven a valuable tool for a wide range of applications. For example, prediction of an election winner, identification of adverse drug reactions, measurement of brand awareness, and many more. However, social media data have not yet been explored for the analysis of the artworld.

This study investigates three fundamental questions: Can artworld actors (artists, museums, galleries, curators, etc.) be identified by mining social media? If so, is it possible to construct an artworld network that visualizes their interactions? Then how an actor's authority on influencing others can be evaluated? Finally, a great challenge is to verify that data available on social media are able to reflect the artworld structure as it exists in the "real" world.

<1>BACKGROUND

The method for identifying artworld actors can be formulated as a community detection problem, a main task of analyzing social media networks. Communities are groups of users that share common properties (e.g., traits, opinions, and preferences). However, the community detection problem is complex and ill-defined since a universal definition of the properties that constitute a community does not exist. Therefore, there are no definite guidelines on how to search for users belonging to a community. An important aspect is that the peculiarities of the network at study should be considered and orient the selected community detection methods for more reliable solutions [4].

Typically, research on community detection initiates by selecting the social media network from which data will be mined. This choice is dependent on the popularity of each social media network among the community at study, the number of publicly available posts, and other characteristics. Subsequently, community detection algorithms can be applied to discover communities in full social networks (e.g., the whole Twitter network) or in smaller topic-specific subsets. In the latter case, it is common for researchers first to devise a methodology for collecting an appropriate dataset, e.g., collect posts that include a specific hashtag.

Social media data are usually modeled as graphs with connected nodes that denote users and their interactions. Such user interactions are: the users mentioned; the posts liked or replied to; the hashtags used; and others. Then, the objective of a community detection algorithm is to find partitions of densely connected nodes, which form a community, and at the same time, are sparsely connected with other communities.

Finally, another challenging research topic in social media network analysis is finding the most influential users. Existing approaches employ a great variety of metrics and criteria to evaluate the influence of users, which are based on the sociology of reputation [5], which first defined the concept of centrality. Centrality is the relative importance of a node within the network to which it belongs.

<1>METHODOLOGY

For our experiments, data will be mined from Twitter [6] that allows users to post short messages, up to 280 characters. Almost every art institution uses Twitter to promote events and, in general, engage with their audience. Other users follow their accounts to stay informed about what is going in the artworld. In contrast, for example, Instagram focuses on generating visual content and its users, in general, tend to browse and like images mostly. Hence, Twitter is appropriate for this study with many publicly available messages of a predominantly textual nature. Nevertheless, the methodology described could be similarly applied to other social media networks.

Community detection methods cannot be utilized per se in the full Twitter network for identifying artworld actors. In such a case, users that interact highly with art-related tweets and have a general interest in art-related processes would be identified. However, this study aims to identify actors who possess certain cultural status and power and thereby define the artworld.

<2>Identifying artworld actors

Defining what an artworld actor is in terms of Twitter interactions, even if we consider it a feasible task, is extremely complicated. For that reason, we pose the question: what if we use a set of exemplar artworld actors that will serve as a seed to identify other artworld actors?

Various published art-related rankings [7] could be utilized to comprise a list of exemplar artworld actors. However, most of them focus on ranking artists. The ArtReview's Power 100 is one of the most well-known rankings that considers not only artists but generally artworld actors. Thus, the Power 100 rankings from 2015 to 2017 are utilized to create a list of Twitter accounts that represent the exemplar artworld actors. Although many criticize the Power 100 for bias in favor of British actors, its continuous publication since 2002 and retaining its legitimacy in the artworld demonstrates that its rankings are plausible enough [8]. Nevertheless, the use of the ArtReview ranking is not hard-wired into the proposed methodology. Many other sources can be tested in future experiments for obtaining a list of exemplar actors.

Finally, the list of exemplars is comprised of 108 unique Twitter accounts representing artworld actors occupying a place in at least one of the three utilized ArtReview rankings. The nationality of exemplars is 31.48% USA, 13.89% UK, and the rest 54.63%, includes actors from various other countries.

All the public tweets posted in 2018 by the exemplar artworld actors are collected, forming the topic-specific subset of Twitter utilized in this study. However, in the following experiments, it will be exemplified that not all exemplar tweets refer to an art-related process. Hence, Natural Language Processing (NLP) methods will be used to detect which tweets refer to art-related events and mentions of artworld actors who do not own a Twitter account.

Following, the analysis of the exemplars' interaction enables the identification of new artworld actors. Specifically, the following interactions are considered: user mentions (the users mentioned in exemplars' tweets); retweets (the users whose posts were retweeted by exemplars); friends (the users that exemplars follow). The above infer an active endorsement of other users as artworld actors. It is regarded that when an exemplar mentions, retweets or follows another user, it contributes to the establishment of this other user as an artworld actor.

<2>Constructing an artworld network

The process for constructing an artworld network is to progressively add nodes connected with directed binary edges [9] that correspond to any of the previously mentioned interactions: user mentions, retweets, and friends. For example, if we consider MoMA as an exemplar artworld actor and retweets a post by MoMA PS1, then the nodes of MoMA and MoMA PS1 will be added in the network connected with a directed edge (Fig. 1). Artforum is mentioned in the same example tweet, and therefore MoMA and Artforum are also connected. Finally, all the users that MoMA follows will be added as nodes in the network connected with it. When this procedure is completed for all exemplars, the result is an artworld network. Finally, the constructed artworld network is analyzed to detect subcommunities [10] of actors among whom many have interacted with each other with the Louvain method [11].

Fig. 1. a) An example tweet by MoMA PS1 that was retweeted by MoMA. b) MoMA is connected with MoMA PS1 because it retweeted one of its posts. MoMA is also connected with Artforum because it is mentioned in the same tweet.

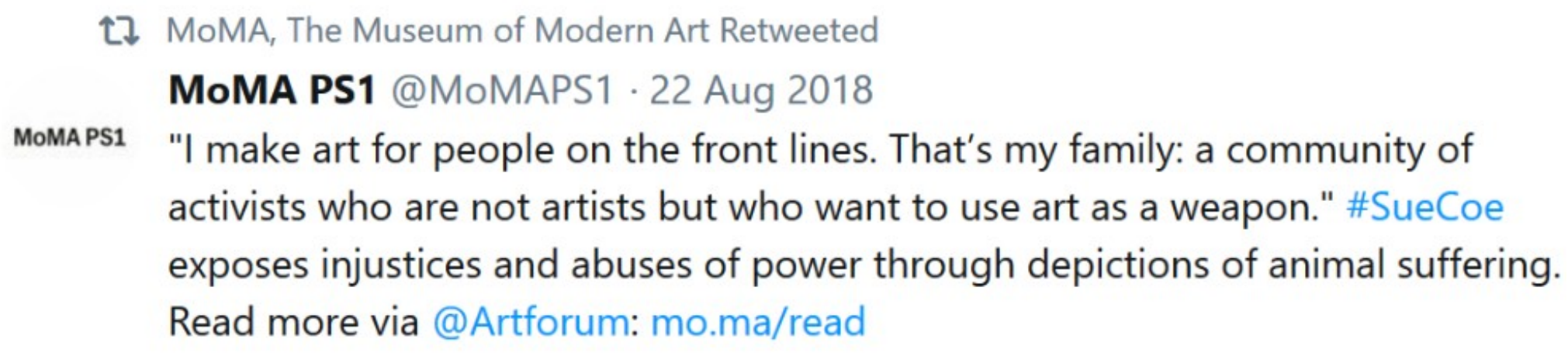

a)

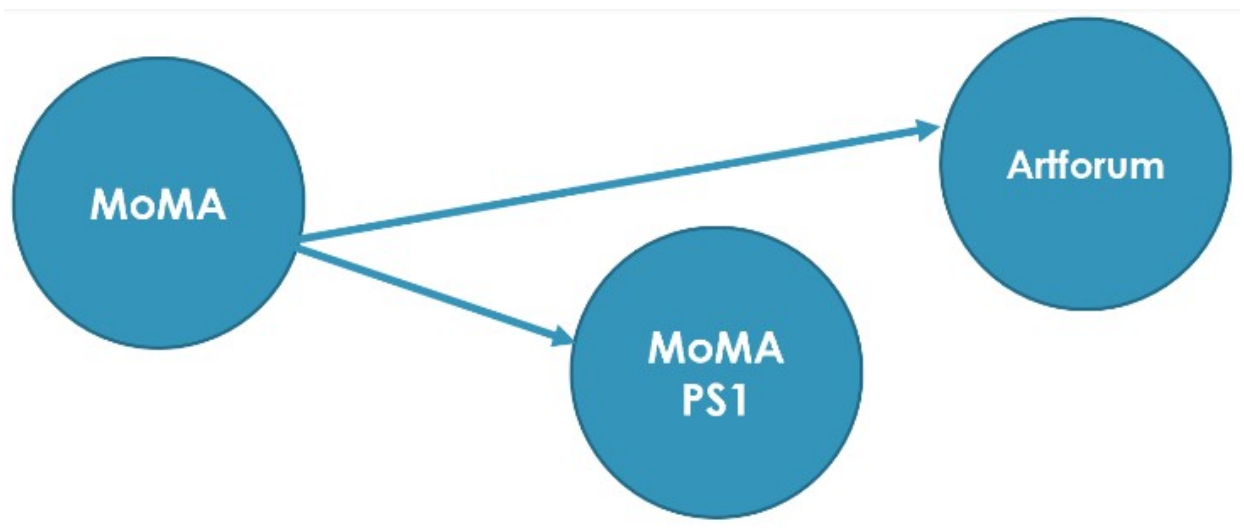

b)

<2>Authority Evaluation

The influence of artworld actors is evaluated with the Hyperlink-Induced Topic Search (HITS) [12] algorithm. The HITS algorithm evaluates iteratively two values for each node: the hub and authority values. Hub values evaluate outward connections, whereas authority values evaluate inward connections. Thus, in the artworld network, a high authority actor is mentioned, retweeted, or followed by many other actors. Accordingly, an actor with a high hub value primarily mentions, retweets, or follows others.

<1>EXPERIMENTS

All the experiments performed are based on the 57,888 collected public tweets posted in 2018 by the 108 exemplar actors [13]. The initial artworld network is comprised only by the exemplar actors (Fig. 2). In all the following figures that depict an artworld network, the color of the nodes represents the subcommunity in which actors belong, and the size of the nodes represents the actors' evaluated authority value.

Fig. 2. The initial artworld network comprised only of the exemplar actors. (© Amalia Foka)

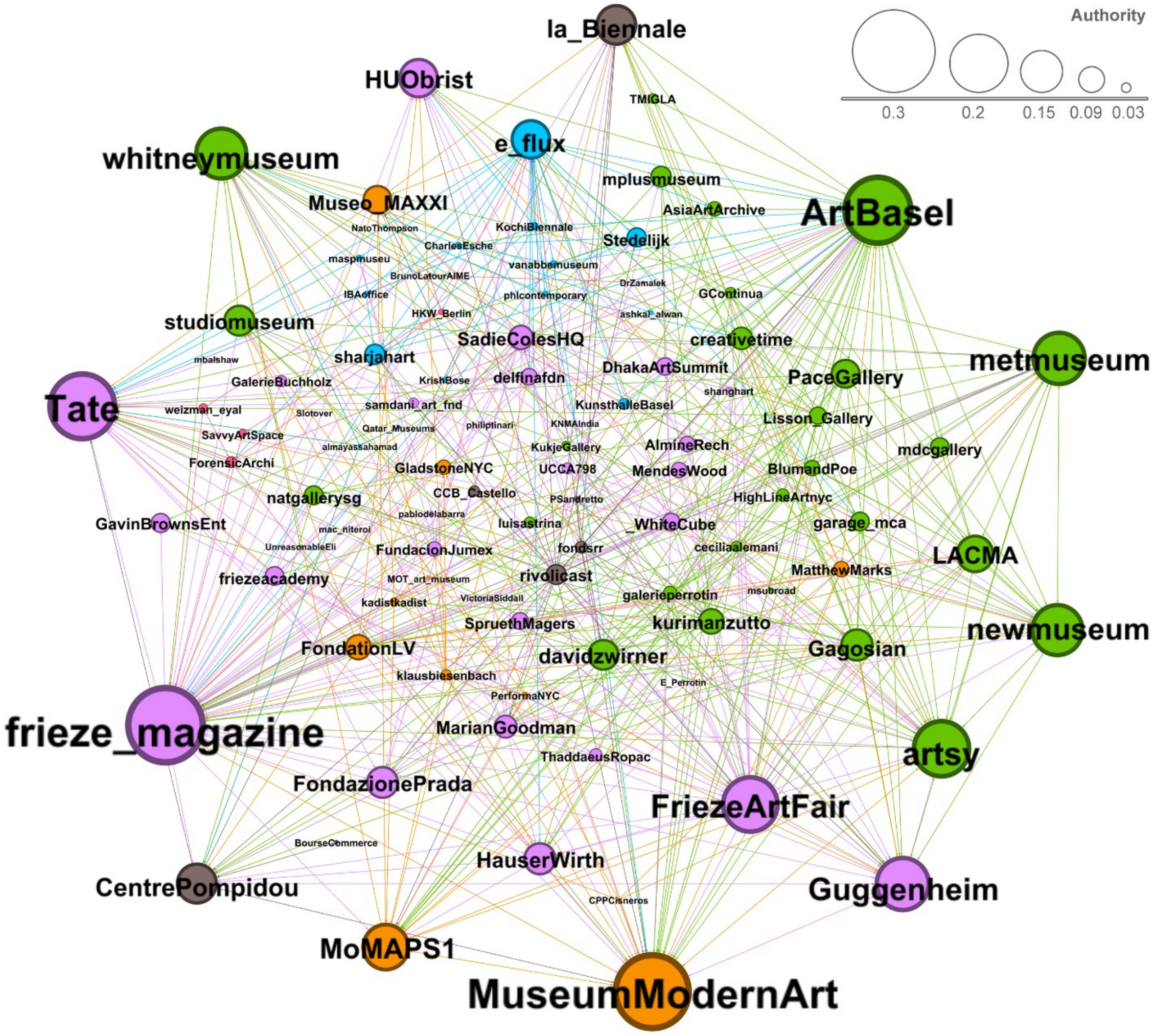

<2>Identifying new artworld actors

Analyzing exemplars' mentions, retweets and friends enables the identification of new artworld actors. More than 30,000 new artworld actors were discovered and included in the artworld network (Fig. 3). There are many new actors with equally high authority as exemplars', such as The Art Newspaper, artnet, Artforum, Hyperallergic, Serpentine Galleries, and the Whitechapel Gallery.

Fig. 3. The artworld network constructed by analyzing all exemplars' interactions. (© Amalia Foka)

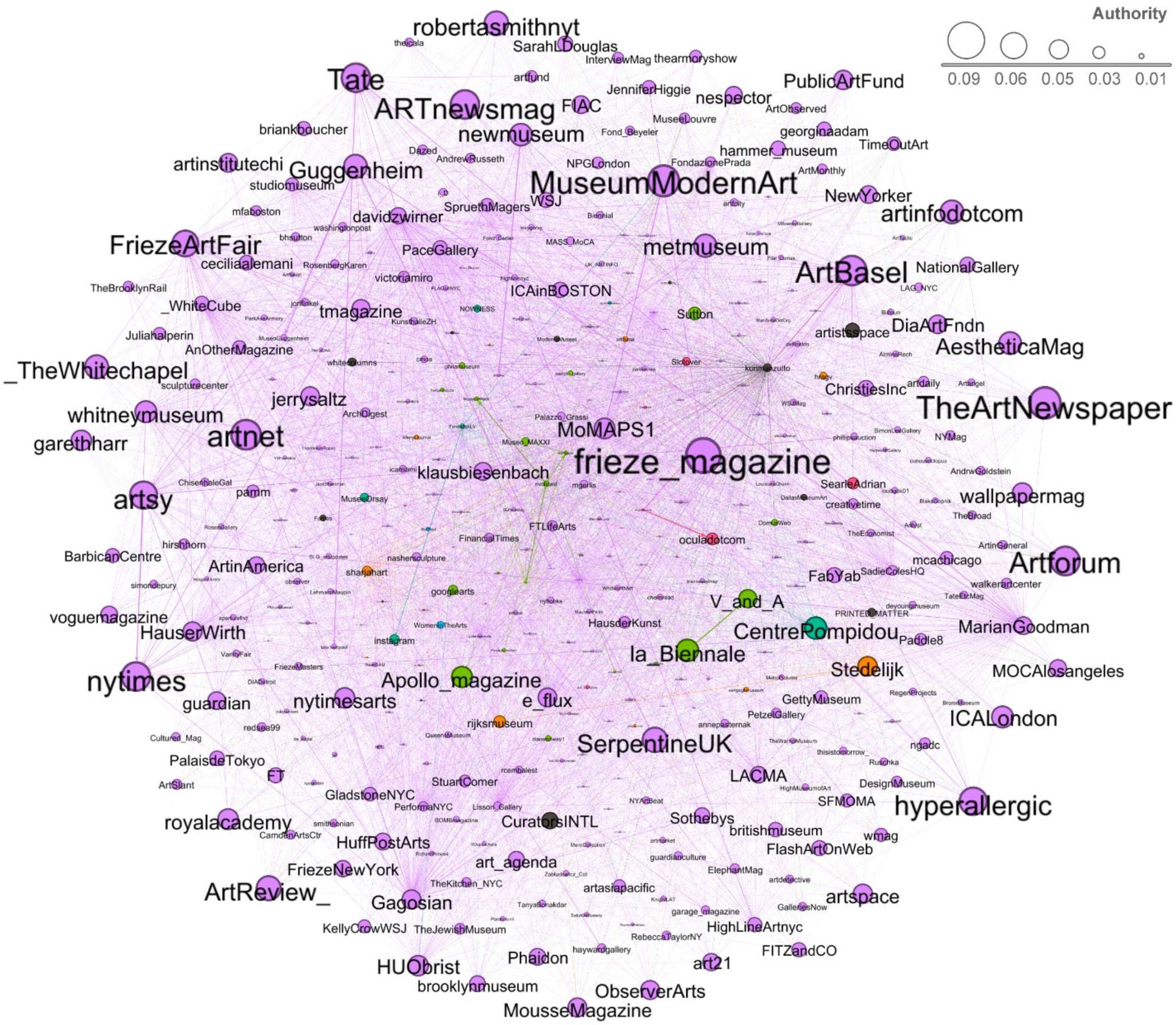

More than half of the highest authority artworld actors have been detected to belong to the same subcommunity. Another seven small subcommunities have been detected. For instance, the subcommunity with Centre Pompidou as its highest authority actor (Fig. 4) represents mainly the actors located in France. Other subcommunities are centered around the actors Biennale di Venezia, Independent Curators International (ICI), National Museum of Women in the Arts, Stedelijk Museum, Ocula Magazine, and Ai Weiwei. A significant disparity in those actors' interactions was found, and hence separate subcommunities have been formed.

Fig. 4. The subcommunity representing mainly the actors located in France. (© Amalia Foka)

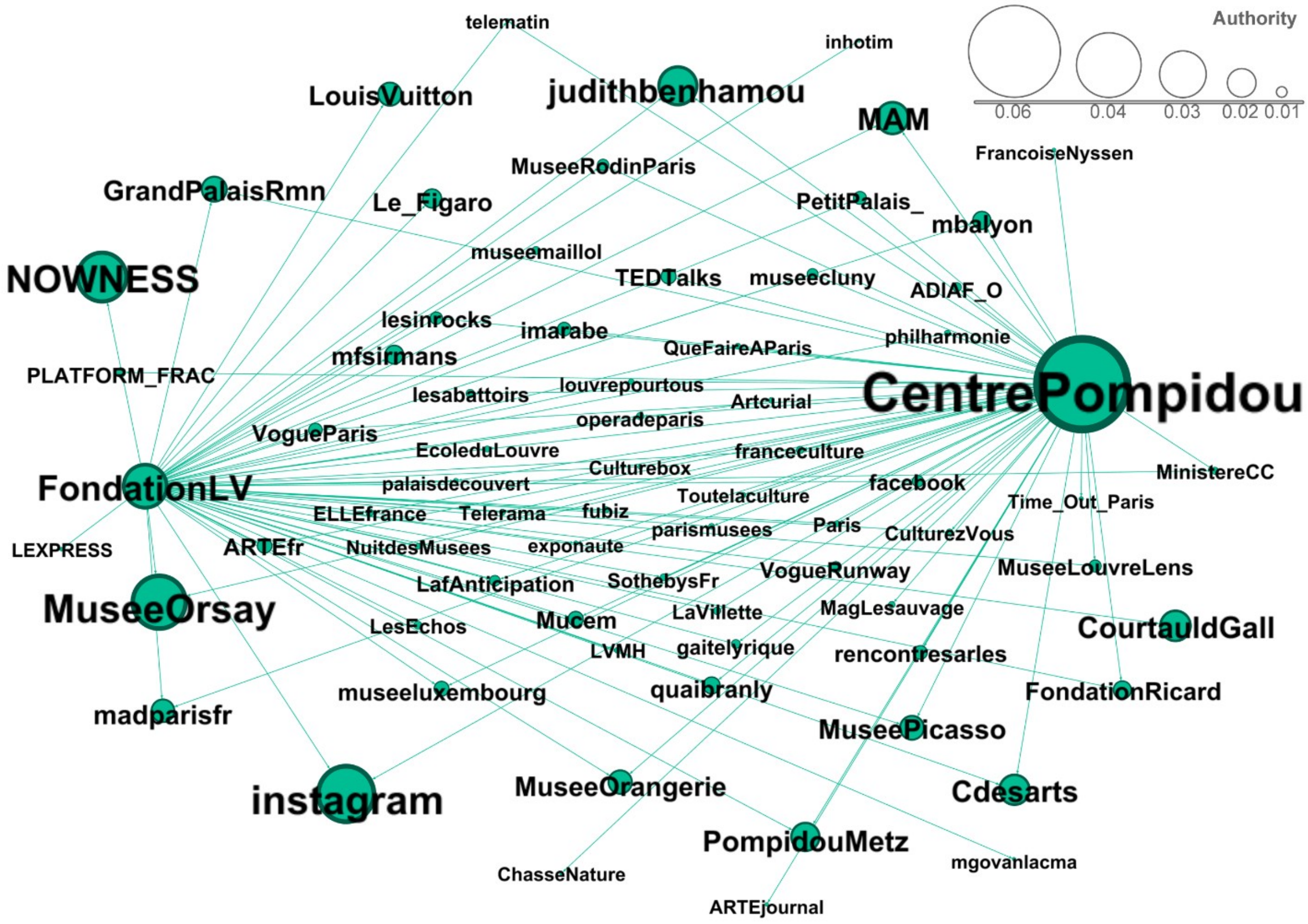

However, there are some nodes included in the network that cannot be regarded as artworld actors. An example is the account of Instagram, mainly mentioned by Fondation Louis Vuitton (Fig. 4). This problem occurs because all interactions are treated equally without examining the actual content of tweets to determine whether they refer to art-related processes. Thus, tweets related to the Instagram photo contest run by Fondation Louis Vuitton were also considered for constructing the artworld network.

<2>Detecting art-related events

In this experiment, only the tweets that refer to art-related events will be analyzed to identify new artworld actors. This approach will treat the issue of including users in the network that cannot be regarded as artworld actors. For that purpose, NLP methodologies are employed to detect exhibitions and other art-related events.

When users post a message that refers to an event, they typically also include the date on which it takes place. However, a date can be written in many different ways, for example, "this Saturday," "next month," "in March." Hence, detecting messages that refer to an event requires resolving temporal expressions, for which TempEx [14] is utilized. Additionally, common temporal expressions in tweets for art-related events like "currently on display," "on show," were used to augment the rules in TempEx.

However, not all tweets that include a temporal expression refer to an art-related process. It is common for exemplars to post tweets that refer, for instance, to the date of birth of an artist. Thus, it is also required to classify tweets to determine if they actually refer to an art-related process. Thus, a probabilistic model, obtained through a semi-supervised training procedure [15], is used for the phrases commonly used for art-related processes. Consequently, when words like "exhibition," "gallery," "curated," "retrospective" are found, tweets are classified as referring to an art-related process. Otherwise, when words like "birthday" or "competition" are found, tweets are classified as non-applicable for the purposes of this study.

After event detection and classification, the size of the initial set of collected tweets posted by exemplars in 2018 reduced from 57,888 to 8,432 tweets. This reduction is not only due to tweets not referring to art-related events but also due to tweets written in languages other than English (estimated at approximately 14,537). Even though non-English tweets are not explicitly excluded, looking for specific English expressions ultimately dismisses them. Thus, it is estimated that about 3,000 tweets referring to art-related events have been lost. Consequently, most tweets from Centre Pompidou, Fundacion Jumex, Kurimanzutto, Museum of Contemporary Art Tokyo (MOT), MAXXI, and a few others were dismissed. Nevertheless, most non-English speaking actors maintained their authority value in large except for Fundacion Jumex and the MOT. Even though many tweets of non-English speaking actors were dismissed, most of them were mentioned by other actors and hence maintained their authority value. Therefore, we can conclude that not considering tweets written in languages other than English does not induce severe issues.

The current artworld network includes 2,188 actors (Fig. 5) and is significantly smaller than the previously presented network (Fig. 3) that included 31,741 actors. Thus, the network constructed by detecting art-related events is far more elitist. Actors like Twitter, Facebook, and Instagram can no longer be found in this network. However, high authority actors like SFMOMA, Whitechapel, Serpentine Galleries, and many more are still included in the current network. When comparing the two networks (Figs 3 and 5), the actors with the most significant increase in their authority value were: National Gallery of Art, Kunstsammlung NRW, Artangel, Printed Matter, Inc., and GalleriesNow.

Fig. 5. The artworld network constructed by considering only tweets that refer to art-related events. (© Amalia Foka)

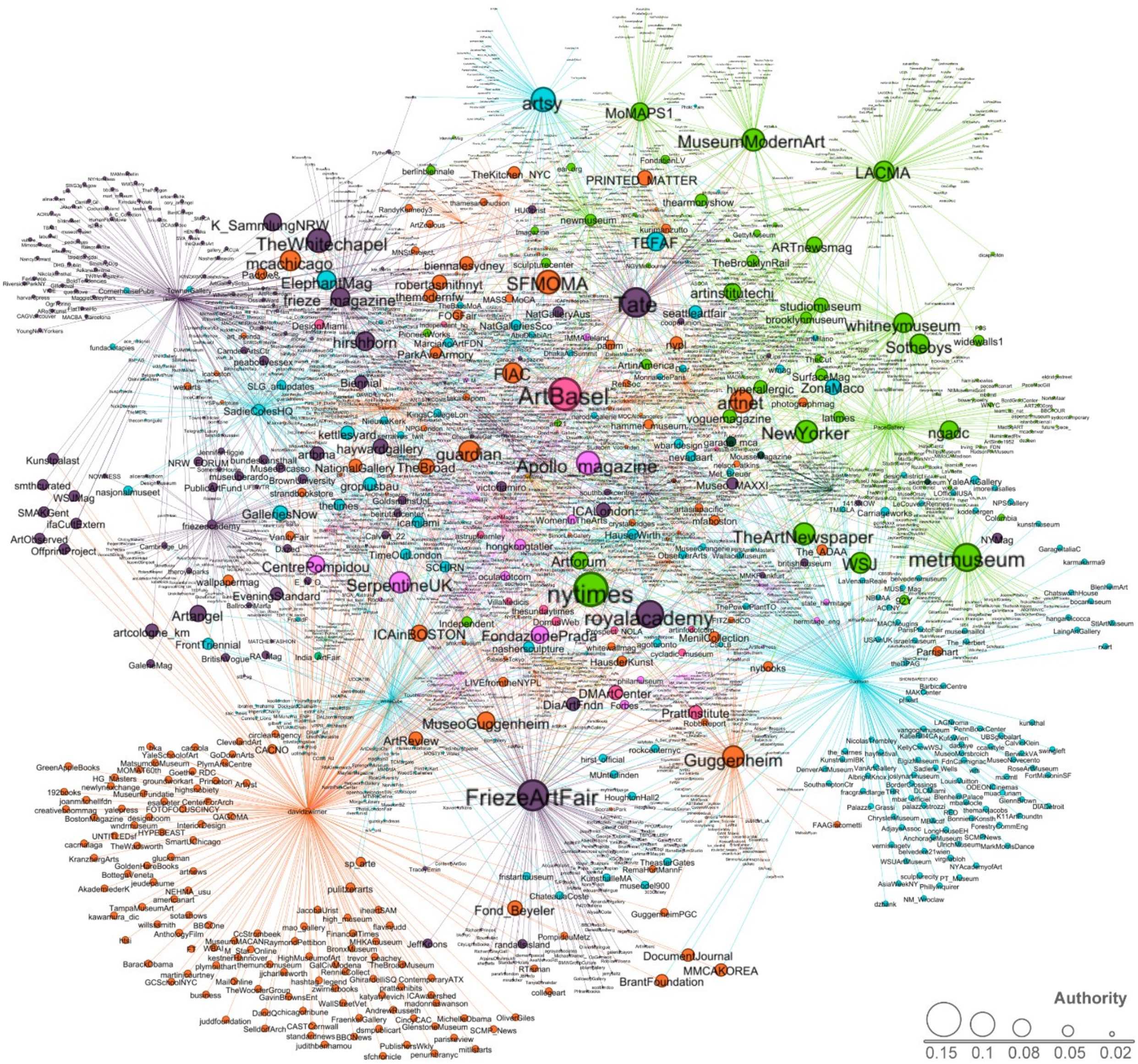

<2>Named Entity Recognition (NER)

The networks presented so far (Figs 3 and 5) primarily include museums, galleries, art-related media, and other organizations. Both networks seem to fail to include artists and persons with other roles in the artworld. The reason for that is that many artworld actors do not own a Twitter account, and even if they do, they are often mentioned by their name rather their Twitter screen name.

To overcome this limitation, Named Entity Recognition (NER) is employed, an NLP task that aims to identify which parts in a text are mentions of entities in the real world. Currently, there are many different approaches for NER that perform well with adequate accuracy. These approaches can identify named entities in a text and classify them as the name of a location, a person, an organization, etc.

In this experiment, a NER approach designed explicitly for tweets, T-NER [16], is applied to recognize the names of artists and other artworld actors that either do not own a Twitter account or they are mentioned by their name. The recognized person and organization names are treated in the same manner as user mentions and therefore added in the network as nodes connected with the actors that mentioned them.

The resulting artworld network consists of 6,862 actors (Fig. 6), including many new actors that do not own a Twitter account and were not identified in the previous experiments [17]. For example, Pablo Picasso, Andy Warhol, Mary Corse, Mark Godfrey, Jackson Pollock, and Franz West. Moreover, quite a few actors' authority values significantly increased, e.g., Victoria Siddall, Yoko Ono, Richard Prince, Gerhard Richter, Andreas Gursky, Julie Mehretu, Tracey Emin, and Hans-Ulrich Obrist. These actors were often mentioned by their name instead of their Twitter screen name, and thus the evaluation of their authority value was not as accurate in the previously presented artworld networks. Finally, it is interesting to note that most commercial art galleries, e.g., Gagosian, have a relatively low authority value but a significantly high hub value.

Fig. 6. The artworld network constructed when Named Entity Recognition is applied. (© Amalia Foka)

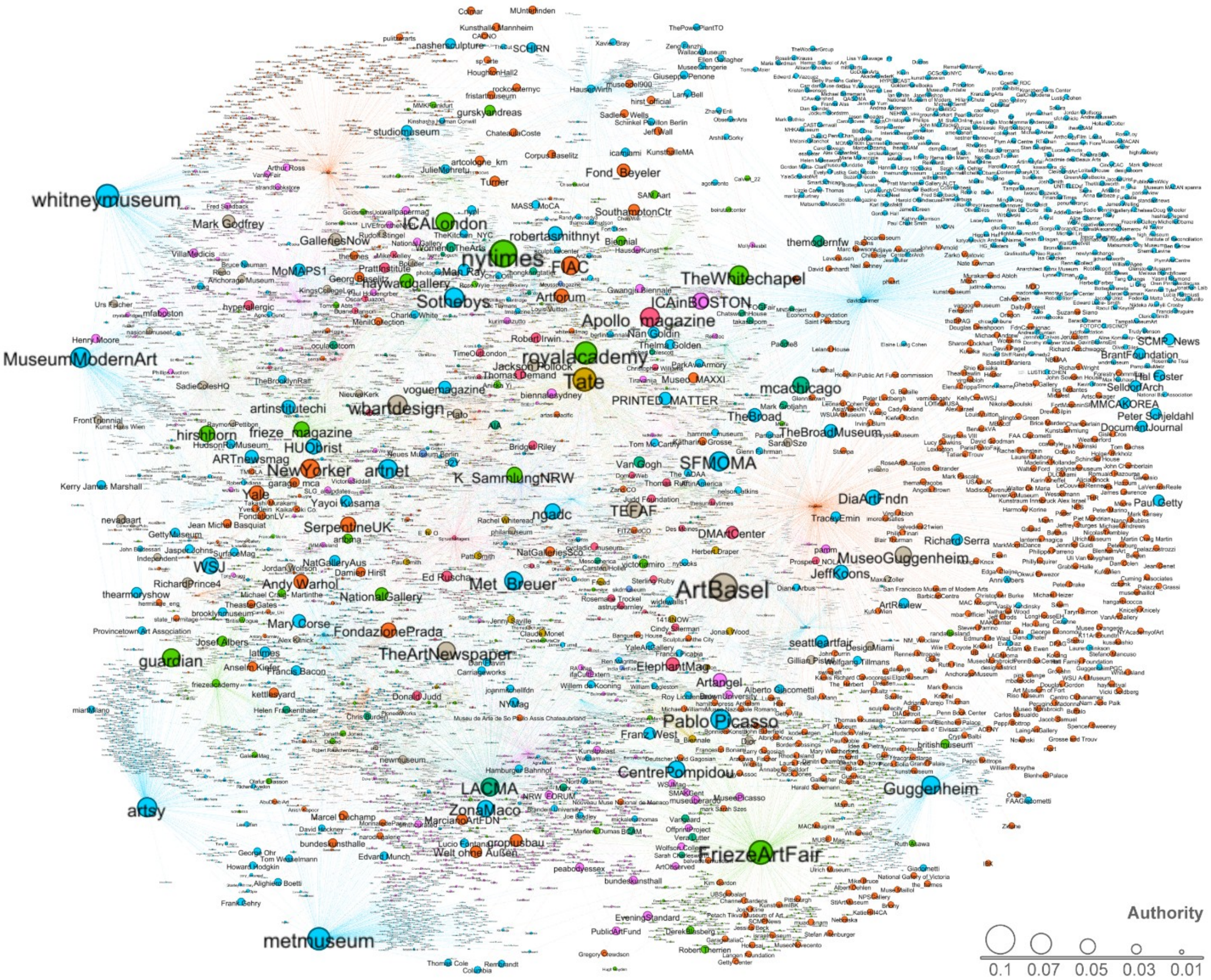

<1>CONCLUSIONS

In this paper, a methodology for constructing a network that maps the current structure of the artworld has been proposed. It has been exhibited that starting with a small set of exemplar actors, numerous new artworld actors and their connections can be identified that resemble the "real" artworld. To be able to explore how social media data can elaborate on the understanding of the artworld, it is a pre-requisite to enable the identification of artworld actors, which has been achieved in this study. Future work includes utilizing various sources for comprising the exemplars list and extending NLP methods for languages other than English to represent the artworld more accurately and eliminate any possible bias. Mining data from multiple social media and thus collecting a more diverse dataset could also contribute in the same direction. Moreover, extracting additional information from social media data can provide more details on artworld actors' interactions, e.g., geolocations, to investigate actors' geographical connections or the specific events that contributed significantly to influence evaluation. Finnaly, executing the proposed methodology in real-time will allow the evaluation of transformations in the artworld as they happen. In conclusion, the proposed methodology can provide a wealth of data that would be very hard to collect otherwise and enable practitioners or theoreticians to insightfully investigate the transformations of the contemporary artworld as well as the role and impact of artworld actors involved in the processes of the production and mediation of art.

<1>REFERENCES AND NOTES

8. A. Quemin, "The Impact of Nationality on the Contemporary Art Market," Sociol. Antropol., **5**, No. 3, 825–-856 (2015).

9. The use of binary edges is justified in Section 6 of the supplemental material.

10. In the supplemental material, a literature review on community detection methods is given in Section 3, whereas alternative network structures are discussed in Section 6.

13. A more detailed discussion of the experiments' methodology and results can be found in the provided supplemental materials.

17. The artworld network can be explored in more detail at http://www.amaliafoka.com/artworld-network/.

<1>BIBLIOGRAPHICAL INFORMATION

Amalia Foka is an assistant professor at the School of Fine Arts, University of Ioannina, Greece. She obtained her Ph.D. (2005) in Computer Science from the University of Crete, Greece. Her research interests include Artificial Intelligence, Robotics, and Computational Creativity.